# Laser-patterned submicron Bi$_2$Se$_3$-WS$_2$ pixels with tunable circular polarization at room temperature

Zachariah Hennighausen[1,*], Darshana Wickramaratne[2], Kathleen M. McCreary[2], Bethany M. Hudak[2], Todd Brintlinger[2], Hsun-Jen Chuang[3], Mehmet A. Noyan[2], Berend T. Jonker[2], Rhonda M. Stroud[2], and Olaf M. van 't Erve[2,*]

[1] NRC Postdoc Residing at the Materials Science and Technology Division, United States Naval Research Laboratory, Washington, D.C. 20375, USA
[2] Materials Science and Technology Division, United States Naval Research Laboratory, Washington, D.C. 20375, USA
[3] Nova Research, Inc., Alexandria, VA 22308, USA

**Abstract**

**Characterizing and manipulating the circular polarization of light is central to numerous emerging technologies, including spintronics and quantum computing. Separately, monolayer tungsten disulfide (WS$_2$) is a versatile material that has demonstrated promise in a variety of applications, including single photon emitters and valleytronics. Here, we demonstrate a method to tune the photoluminescence (PL) intensity (factor of x161), peak position (38.4meV range), circular polarization (39.4% range), and valley polarization of a Bi$_2$Se$_3$-WS$_2$ 2D heterostructure using a low-power laser (0.762μW) in ambient. Changes are spatially confined to the laser spot, enabling submicron (814nm) features, and are long-term stable (>334 days). PL and valley polarization changes can be controllably reversed through laser exposure in vacuum, allowing the material to be erased and reused. Atmospheric experiments and first-principles calculations indicate oxygen diffusion modulates the exciton radiative vs. non-radiative recombination pathways, where oxygen absorption leads to brightening, and desorption to darkening.**

* Authors for correspondence, E-mail:  zachariah.hennighausen.ctr@mail.nrl.navy.mil; olaf.vanterve@nrl.navy.mil;

**Introduction**

Circularly polarized light (CPL) interacts differently with materials than linearly polarized light, providing a separate avenue to probe and manipulate materials. This property has facilitated a variety of existing and emerging technologies, including bioinspired navigation,[1] spintronics,[2] quantum computing,[3,4] quantum communication,[5,6] non-linear image encryption,[7] valleytronics,[8] and non-invasive cancer screening.[9] Manipulating circular polarization traditionally requires bulky optical elements and intricate setups, which limits its application and greater adoption. Recent advances in nanophotonics,[10] organic chiral molecules,[11] and metasurfaces[12] have produced impressive results, but such nanoscale fabrication can be challenging, and often the properties cannot be modified after fabrication. In this work, we demonstrate technology where the circular polarization can be modified across a large range, at room temperature, using a low-power laser, and with submicron resolution. The changes can be reversed using a laser in vacuum, enabling write-read-erase-reuse applications. Amongst other applications, the findings pave the way for laser-patterned, submicron circular polarization sensors, pixels, and optical elements.

Two-dimensional (2D) transition metal dichalcogenides (TMDs) - notably $MoSe_2$, $MoS_2$, $WSe_2$, and $WS_2$ - have demonstrated considerable promise for CPL applications. Additionally, the materials can be grown on an economical scale using chemical vapor deposition (CVD), providing a facile route toward application. Using CVD, we first grow monolayer $WS_2$ and then few-layer $Bi_2Se_3$ on top. We use $Bi_2Se_3$ as a medium to manipulate the optical, exciton, and valley properties of $WS_2$.

Monolayer $WS_2$ is a direct band gap semiconductor with $\sigma+$ and $\sigma-$ transitions at the *K* and *K'* valleys, respectively, due to time-reversal symmetry and a strong spin-orbit coupling.[13,14] When exposed to $\sigma+$ ($\sigma-$) CPL with energies above the band gap, tightly-bound excitons and trions (i.e., electron-hole quasiparticles) form in the *K* (*K'*) valley. When these excitons recombine, CPL is emitted with a circular polarization corresponding to the valley. The valley-dependent selection rules allow the *K* and *K'* valleys to be selectively populated and manipulated using only

CPL, suggesting $WS_2$ is ideal for advanced information processing schemes and other valleytronics applications.[8]

Excitons formed in TMDs can scatter between the valleys and recombine with the opposite CPL, which reduces the degree of circular polarization (DoCP) of the material. Several intervalley scattering mechanisms have been demonstrated (See Section S7 for additional information). Phonon-assisted scattering[15–18] and the Maialle-Silva-Sham (MSS) exchange interaction[16,19–21] are considered to be the most dominant mechanisms in monolayer TMDs, providing a framework to engineer better materials. A variety of factors influence a TMD's DoCP, including defects,[22] exciton lifetime,[23–26] dopants,[27] temperature,[28] magnetic fields,[29] excitation wavelength,[30] and carrier density.[31] Despite these advances, manipulating the material deterministically *in situ* with submicron resolution remains a challenge. Using a back-gated device, the carrier density and DoCP can be tuned real-time, but the effect is not stable and disappears when power is removed.[26,31] A laser can be used to dope $H_2$-activated monolayer $MoS_2$ with spatial-selectivity and manipulate the DoCP, but it was only demonstrated at very low-temperatures (4K) and the changes are not reversible.[27]

In this work, we demonstrate a method to manipulate the valley polarization and DoCP across a large 39.4% range, using a low-power laser (0.762μW) in ambient and at room temperature. Additionally, the photoluminescence (PL) intensity and peak position can be tuned with high precision across a factor of x161, and 38.4meV range, respectively. The changes are locally confined, enabling features as small as 814nm, and are long-term stable (>334 days). Applying a laser in vacuum controllably reverses changes to the PL and DoCP. The PL and DoCP can be reliably manipulated within a range, allowing the material to be erased and reused. Atmospheric measurements, Fick's 2nd Law of Diffusion fitting, and first-principles calculations indicate oxygen absorption and desorption, facilitated by the few-layer $Bi_2Se_3$, is responsible for the large modulation of the exciton radiative vs. non-radiative recombination pathways. We propose a framework to understand the changes in PL and valley polarization due to oxygen absorption and desorption. Our work demonstrates precision tuning of spatially-selective submicron pixels furthering numerous technologies, including multistate multidimensional optical data storage, valleytronics, and quantum information sciences.

## Experiment and Results

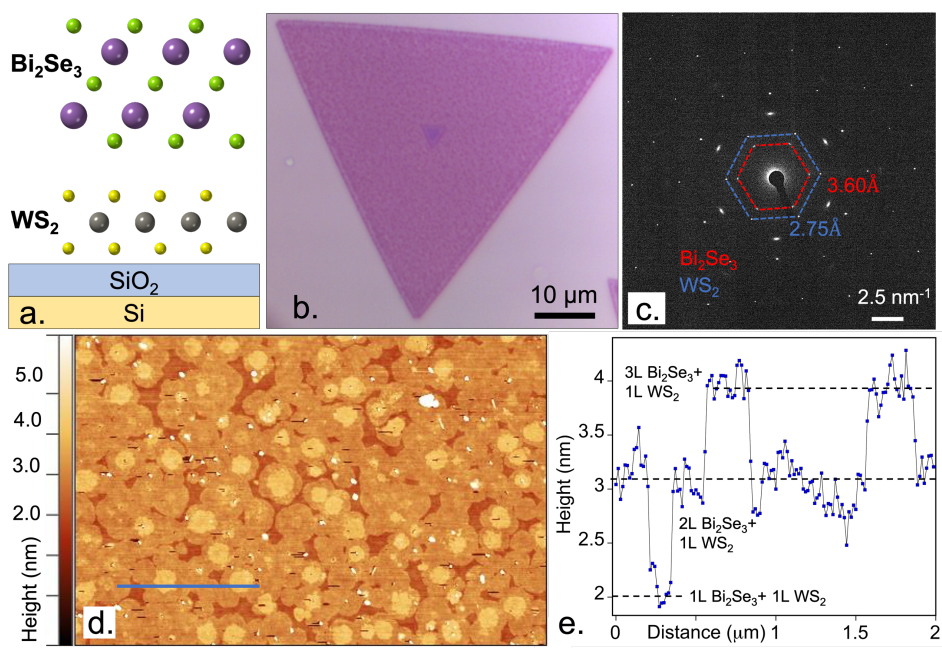

**Figure 1: As-grown Bi$_2$Se$_3$-WS$_2$ 2D heterostructure.** (a) Schematic of as-grown Bi$_2$Se$_3$-WS$_2$ 2D heterostructure. (b) Optical microscope image. (c) TEM diffraction pattern showing crystalline Bi$_2$Se$_3$ and WS$_2$. The majority of the Bi$_2$Se$_3$ is oriented within ±2° of the 0° twist angle. (d) AFM scan of sample in (b), showing Bi$_2$Se$_3$ grows uniformly. (e) Height line profile along the blue line inset in (d). Bi$_2$Se$_3$ grew mostly bilayer, with gaps of monolayer, and islands of trilayer.

Figure 1 shows the characterization of a representative as-grown Bi$_2$Se$_3$-WS$_2$ 2D heterostructure. Monolayer WS$_2$ is grown on an Si/SiO$_2$ substrate, and then 1-3 quintuple layers of Bi$_2$Se$_3$ are grown on top of the WS$_2$. Both growths use chemical vapor deposition (CVD), suggesting the method is economically scalable, providing a route to application. Figure 1a shows a schematic of the 2D heterostructure, and Figure 1b shows an optical image of a representative sample.

Figure 1c shows a selected area electron diffraction (SAED) pattern from an ~ 200 nm region. The majority of the Bi$_2$Se$_3$ is oriented within ±2° of the 0° twist angle, although some crystallites with disorientations up to 10° were observed, in line with previous work.[32] While WS$_2$ was crystalline throughout, Bi$_2$Se$_3$ was found to have nanoscale areas of high crystallinity, surrounded by a discontinuous Bi$_x$Se$_y$ alloy that showed no clear structure. We suspect the few-layer Bi$_2$Se$_3$ becomes unstable and is altered during transfer from its initial SiO$_2$ growth substrate onto TEM-suitable support.

Figure 1d is a representative atomic force microscope (AFM) scan, and Figure 1e is the height profile along the blue line inset. The $Bi_2Se_3$ grew primarily as bilayers, with regions of monolayers and islands of trilayers (Figure S1 contains amplifying AFM data and discussion). The $Bi_2Se_3$ exhibits a pronounced Raman signal typical of few layer $Bi_2Se_3$ (Figure 2),[33,34] and it grows as well-defined hexagons (Figure S2), suggesting it grows with long-range crystallinity. A majority of the hexagons' sharp edges are qualitatively parallel to that of $WS_2$, in agreement with TEM measurements.

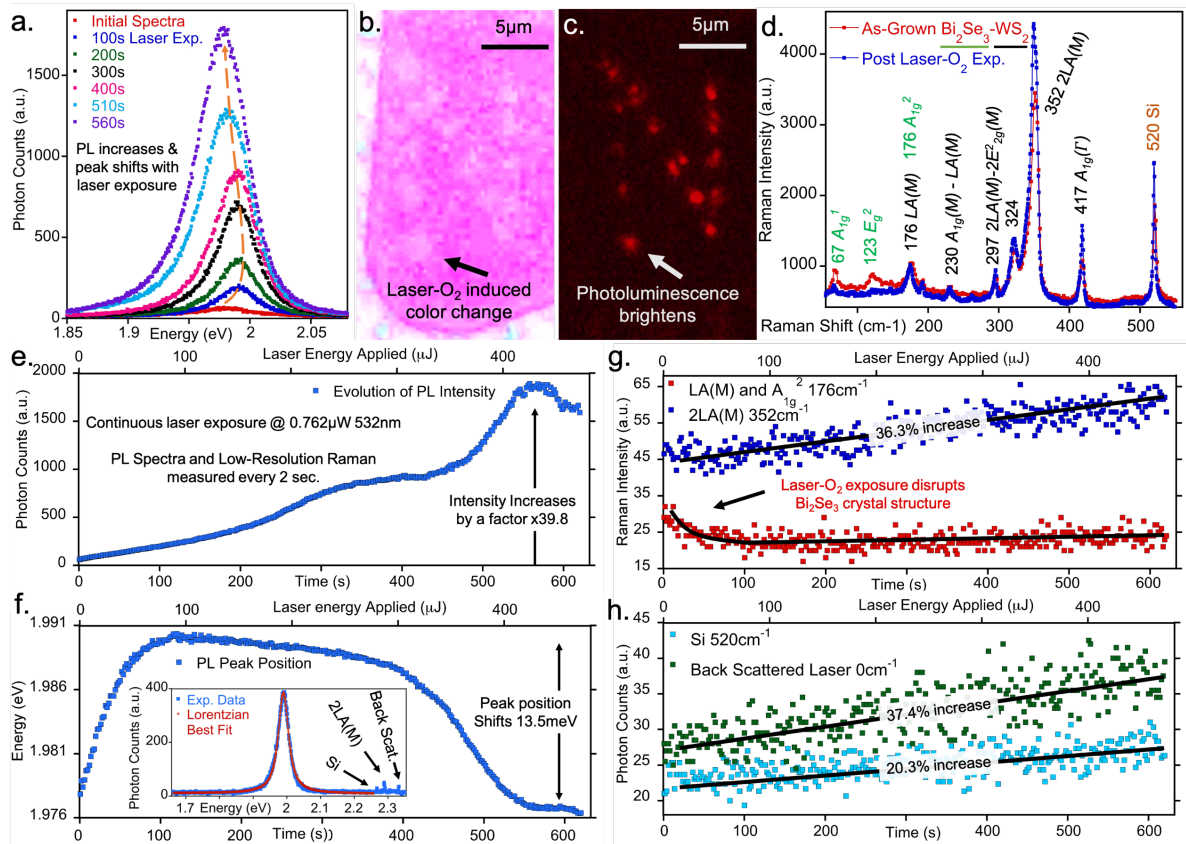

**Figure 2: Low-power laser exposure in ambient modifies the Photoluminescence and Raman spectra.** (a) Various PL spectra after different laser exposure times in ambient. The PL intensity increases and peak position shifts with laser-oxygen exposure. (b) Optical microscope and (c) fluorescence images of a sample with the letter "N" patterned. The optical image shows a color change from purple to white, and the fluorescence image demonstrates the ability to write submicron (814nm) pixels. (d) Raman spectra before and after laser exposure, where the $Bi_2Se_3$ and $WS_2$ peaks are labeled with green and black, respectively. $Bi_2Se_3$ peaks decline, likely due to degradation of the crystal structure, while all other peaks increase. (e)-(f) Evolution of (e) PL intensity and (f) peak position as a function of uninterrupted low-power (0.762μW) laser exposure with a 50x objective (0.050μW/μm²). Inset: PL spectrum after 200s laser exposure with a Lorentzian fit, and the concurrent low-resolution Raman spectrum. (g)-(h) Evolution of three Raman peaks and the backscattered laser peak as a function of laser exposure. The peaks were collected concurrent with the PL spectra using a low-resolution grating (150g/mm), allowing data in (e)-(h) to be directly compared. (g) The peak centered at 176cm$^{-1}$ is a combination of both the $Bi_2Se_3$ ($A_{1g}^2$) and a $WS_2$ (LA(M)) Raman peak. (h) The silicon peak increase suggests that the transmission increases. The backscattered peak increases faster than the silicon peak, suggesting the reflection increases.

Figure 2 demonstrates low-power laser exposure in ambient steadily modifies the PL and Raman spectra of a $Bi_2Se_3$-$WS_2$ 2D heterostructure, where the changes are confined to the laser spot. Figure 2a shows the PL spectra at various times after uninterrupted exposure to a 0.762μW 532nm laser with a 50x objective (0.050μW/μm$^2$). Initially, the PL intensity is quenched; however, as time progresses, the intensity increases and peak position shifts to higher energies, before reversing direction and shifting to lower energies. Figure 2b-c are optical and fluorescence images, respectively, of a 2D heterostructure with the letter "N" patterned. Laser exposure in an atmosphere with oxygen induces a color change from purple to a variant of white. The fluorescence image demonstrates submicron (814nm) feature resolution and pixel size. As discussed later, oxygen is required concurrent with laser exposure to induce changes.

Figure 2d shows the Raman spectrum of a $Bi_2Se_3$-$WS_2$ 2D heterostructure before and after laser exposure in ambient, where green and black writing correspond to $Bi_2Se_3$ and $WS_2$, respectively. The well-formed peaks suggest both $Bi_2Se_3$ and $WS_2$ are crystalline, in agreement with TEM diffraction measurements (Figure 1c). All the peaks match previously published literature for monolayer $WS_2$ and few-layer $Bi_2Se_3$.[33,34] Of note, $Bi_2Se_3$ Raman modes evolve as the material is thinned from bulk to monolayer. Our data suggests the $Bi_2Se_3$ is between 1-4 layers, in agreement with AFM measurements (Figure 1d). After laser exposure in an atmosphere with oxygen, the $Bi_2Se_3$ Raman peaks decrease intensity, eventually disappearing, suggesting the crystal structure is being altered. This is in agreement with previous work that showed laser exposure in ambient of $Bi_2Se_3$-TMD 2D heterostructures disrupts the $Bi_2Se_3$ crystal structure.[35,36] Conversely, the $WS_2$ Raman peaks, silicon Raman peak, and backscattered laser all increase intensity.

Figure 2e-f show the evolution of the PL intensity and peak position, respectively, as a function of uninterrupted low-power (0.762μW) 532nm laser exposure with a 50x objective (0.050μW/μm$^2$) in ambient. A spectrum was collected every two seconds, for 620 seconds, demonstrating the material's smooth and controlled response with very low noise. The intensity increases a factor x39.8, while the peak position shifts 13.5meV. The line-shape contains multiple non-linear features, such as the plateau in the middle, that likely are due to the Bi$_2$Se$_3$ crystal being altered. The PL intensity and peak position data were extracted from each spectrum by fitting with a Lorentzian function and linear background. A robust fitting algorithm written in Python Spyder software ensured the global minimum was found with low uncertainty.[37] The good fit suggests only one quasiparticle recombination dominates the PL, and that the model captures the system well (see inset and Section S2 for representative fits). The inset shows the PL spectrum at 200s with a Lorentzian fit, and the low-resolution Raman spectrum measured concurrently. The spectra in Figure 2a correspond to the data in Figure 2e-f.

Figure 2g-h show the evolution of three Raman peaks and the backscattered laser peak as a function of laser exposure in ambient. The peaks were collected concurrent with the PL spectra in Figure 2e-f using a low-resolution grating (150g/mm), allowing the data to be directly compared. Figure 2g shows the Bi$_2$Se$_3$ (A$_{1g}^2$) and WS$_2$ LA(M) peaks centered at 176cm$^{-1}$, and the WS$_2$ 2LA(M) peak at 352cm$^{-1}$. The 176cm$^{-1}$ peak initially declines, likely due to the Bi$_2$Se$_3$ (A$_{1g}^2$) peak decreasing, suggesting the Bi$_2$Se$_3$ crystal is being altered. The LA(M) peak increases on average 36.3% with laser exposure, in agreement with Figure 2d. Figure 2h shows the silicon Raman peak and the backscattered laser increase 20.3% and 37.4%, respectively. The silicon peak increase suggests that the Bi$_2$Se$_3$-WS$_2$ transmission increases. The backscattered peak is a measure of both the reflection and transmission because a portion of the signal is reflected from the silicon. The backscattered peak increases faster than the silicon peak, suggesting the reflection increases. Together, this data offers a possible explanation why the WS$_2$ Raman peaks increase with laser exposure (Figure 2d,g): As Bi$_2$Se$_3$ becomes more transparent, more light enters WS$_2$.

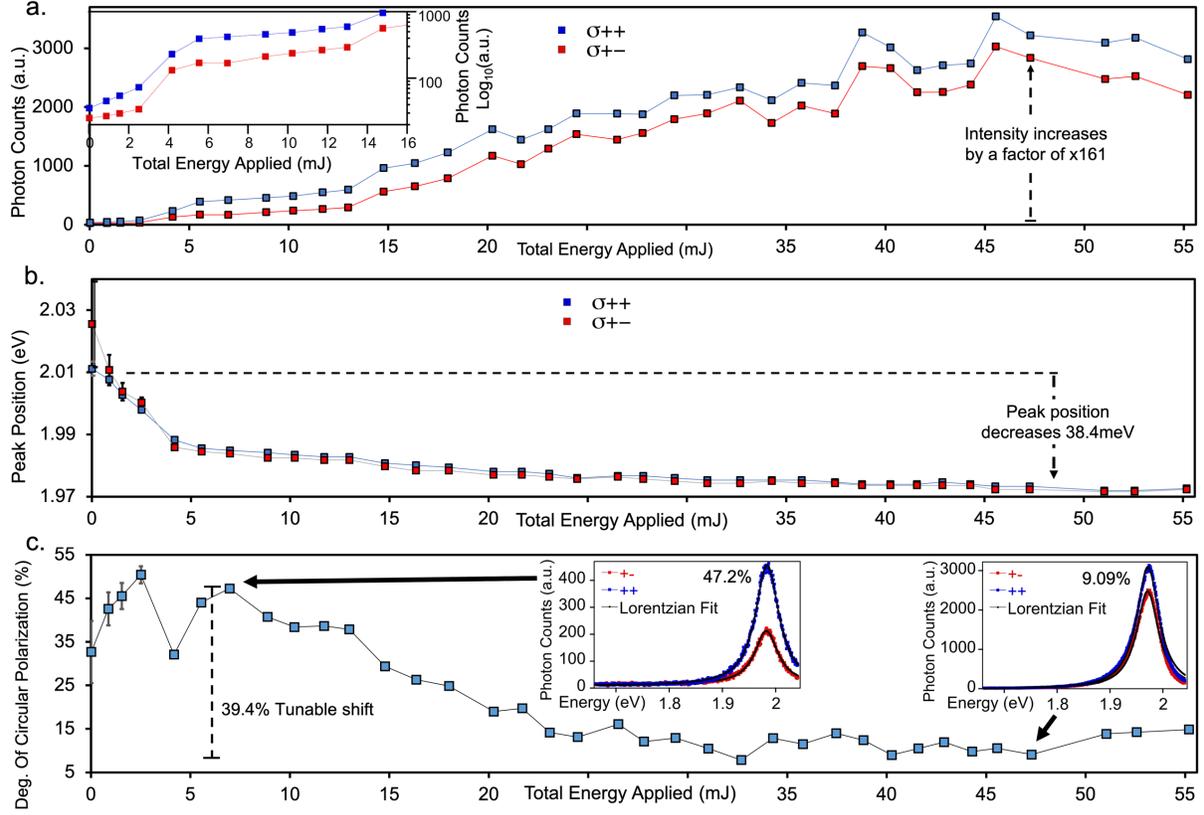

**Figure 3: Tunable circular polarization using a laser in ambient at room temperature.** (a)-(c) Evolution of PL (a) intensity, (b) peak position, and (c) DoCP (valley polarization) subject to intermittent exposure of a 13.5μW σ+ circular polarized 588nm laser with a 100x objective (1.25μW/μm²). (c) The circular polarization and valley polarization shifts 39.4%, demonstrating the large range the material can be manipulated. Insets: The σ++ and σ+- spectra with Lorentzian fits for the 7th and 33rd data points, demonstrating the robust fitting captures the spectra. Error bars are plotted on all data points, showing 2-sigma (95.4%) confidence intervals.

Figure 3 demonstrates tunable DoCP and valley polarization using intermittent exposure of a 13.5μW positive helicity (σ+) circular polarized 588nm laser with a 100x objective (1.25μW/μm²). Data was collected for both positive (σ++) and negative (σ+-) helicity. The DoCP is calculated using Eq. (1).

$$P_{Circ} = \frac{I(\sigma_+) - I(\sigma_-)}{I(\sigma_+) + I(\sigma_-)} \qquad (1)$$

Figure 3a-b shows the evolution of the PL intensity and peak position, respectively, for the σ++ and σ+- states. The intensity increases over a factor of ×161, while the peak position decreases 38.4meV. The DoCP (Figure 3c) initially increases before steadily decreasing and plateauing. We highlight that the PL behavior was reproduced over multiple trials on different samples. Section S2 contains representative Lorentzian fitting of the σ++ and σ+- spectra and amplifying analysis.

All measurements were done in ambient at room temperature, and changes can be reversed and reliably manipulated within a range (discussed later), making the technology attractive for applications. To the best of our knowledge, we demonstrate one of the largest ranges of *in situ*, spatially-selective, and tunable DoCP at room temperature, where changes can be reversed allowing the material to be erased and reused.

Error bars are calculated by finding the 2-sigma (95.4% confidence) value around the chi-squared minima for the applicable variables, and propagating the error as appropriate.[37] Note, as the spectra intensity and signal-to-noise ratio increase, the uncertainty decreases, making error bars difficult to distinguish on the graph.

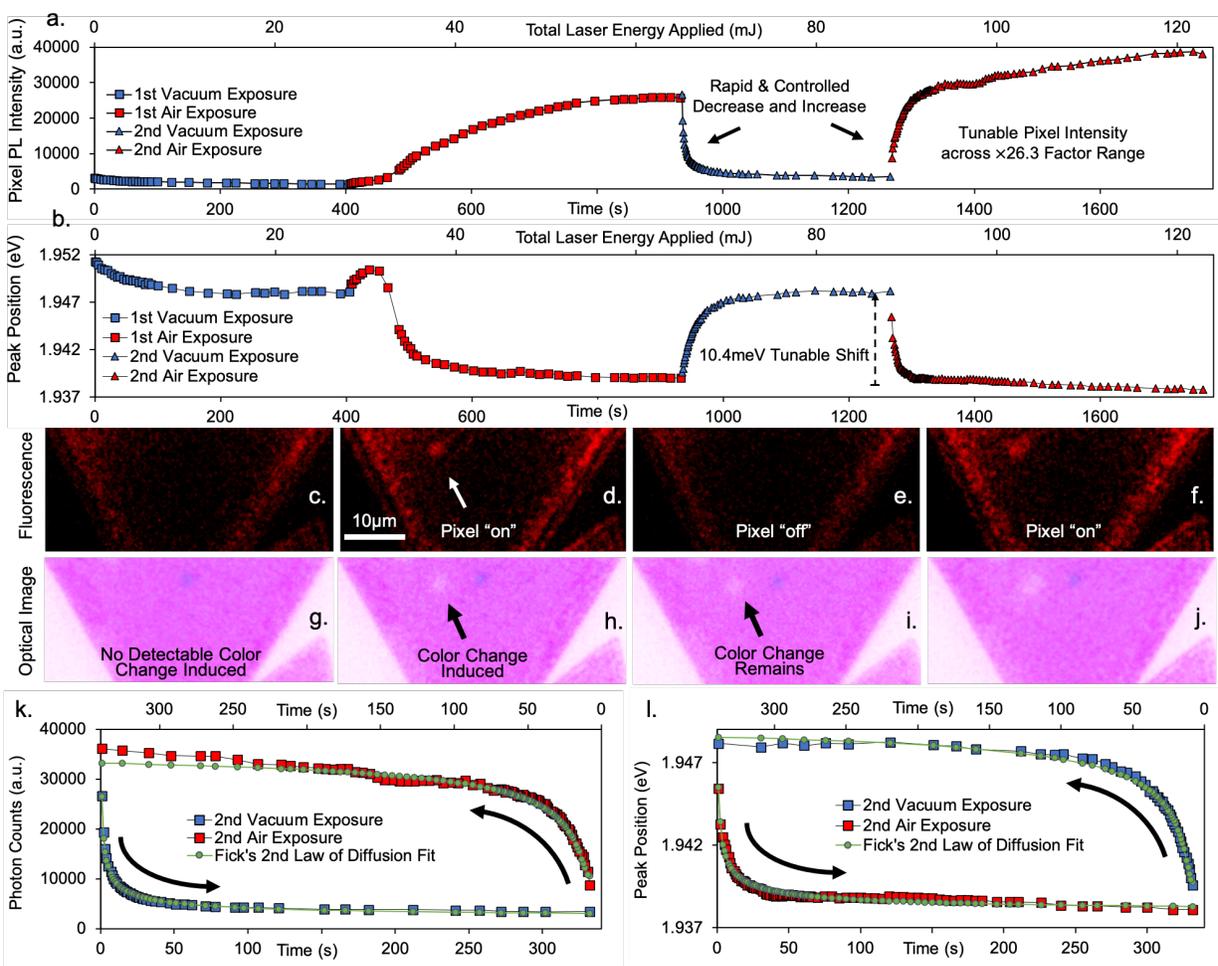

**Figure 4. Rewritable pixels with hysteretic behavior through controlled oxygen absorption and desorption.** (a)-(b) Evolution of (a) PL intensity and (b) PL peak position as a function of laser exposure in air (760 Torr) and vacuum (0.226 Torr) environments. (c-f)

Fluorescence images and (g-j) optical images corresponding to the data in (a) and (b). Fluorescence images demonstrate how the pixel brightness can be modulated, and state flipped between "on" and "off". (c) Post 1st vacuum. (d) Post 1st air. (e) Post 2nd vacuum. (f) Post 2nd air. (g) No color changed induced by the laser. (h) Color change induced. (i), (j) Color change remains. (k)-(l) The PL Intensity and peak position from 2nd vacuum exposure and 2nd air exposure shown in (a) and (b) are replotted. The close fit with Fick's 2nd Law of Diffusion suggests diffusion into and out of the material is present.

Laser exposure in vacuum reverses local changes to the PL intensity and peak position (Figure 4). Our findings suggest the PL is manipulated through oxygen diffusion into and out of the 2D heterostructure. The sample was exposed without interruption to a 38.8μW 532nm laser at a 50x long working distance objective (1.65μW/μm$^2$).

Figure 4a-b show the evolution of the PL intensity and peak position, respectively, as a function of two atmospheres (i.e., air and vacuum), and laser exposure. When the laser is initially applied in vacuum, the PL intensity decreases 54.1% and the peak position shifts slightly (3meV) lower in energy. When the environment is switched to air, the PL increases by a factor ×16.9, and the peak position initially shifts slightly upward (2.3meV), before shifting lower in energy (11.5meV). When the environment is switched back to vacuum (2nd round), the PL intensity decreases 86.5% and the peak position shifts 9.4meV upward. When the environment is switched back to air (2nd round), the PL intensity increases a factor ×11.1 and the peak position decreases 10.4meV.

The fluorescence and optical images shown in Figure 4c-f and Figure 4g-j, respectively, were taken after the last data point in each atmosphere. The fluorescence images demonstrate a pixel's state can be flipped between "on" and "off", and its brightness modulated with spatial selectivity. The optical images show the characteristic color change appears after the first laser exposure in air, but does *not* reverse through laser exposure in vacuum. Similarly, changes to the Raman spectra are *not* readily reversed (Figure S5). When exposed to a laser in air, numerous peaks increase intensity, but do not reverse in vacuum. These findings suggest the material undergoes an irreversible, structural transformation of the $Bi_2Se_3$ crystal structure during the first laser-oxygen exposure. Previous work using $Bi_2Se_3$-$MoS_2$[35] and $Bi_2Se_3$-$MoSe_2$[36] 2D heterostructures demonstrated that laser-oxygen induced optical and Raman changes could be reversed through >3 hours of annealing in an argon environment, possibly because the $Bi_2Se_3$ was recrystallized.

As shown in Figure 4a-b, the first laser-oxygen exposure has a different curve shape compared to subsequent exposures, possibly because the $Bi_2Se_3$ crystal is being altered during exposure. Notably, the first laser-oxygen exposure evolves slower and contains numerous non-linear features that are not present in subsequent exposures. For example, the peak position increases before decreasing, and the intensity curve plateaus before increasing again. This behavior is repeated across numerous samples (Figure 2, Figure S6, Figure S7, and Figure S12). Conversely, subsequent exposures (i.e., 2nd vacuum and 2nd Air) evolve much faster, and have a recurring "logarithmic" curve shape that suggests the response is deterministic, and the underlying mechanisms facilitates repeatable behavior.

To elucidate the mechanism behind the 2nd vacuum and 2nd air curves, we fit the data with Fick's 2nd Law of Diffusion (Figure 4k-l).[38] Fick's 2nd law predicts how the concentration changes with respect to time. Assuming fixed concentration boundary conditions (i.e., air, and the $SiO_2$ substrate), Fick's 2nd law reduces to Eq. (2), where $c$ is the concentration, $x$ is the distance from the 2D heterostructure surface, $D$ is the diffusion coefficient, and $t$ is the time. Since a majority of the sample probed had two quintuple layers of $Bi_2Se_3$, we set x = 2nm,

$$c(x,t) = c_{air} - (c_{air} - c_{initial})\frac{2}{\sqrt{\pi}}\int_0^z e^{-y^2}dy; \quad z = \frac{x}{2\sqrt{Dt}} \qquad (2)$$

The close fit of the 2nd vacuum and 2nd air curves suggests that diffusion is central to the mechanism (Figure 4k-l). Air is approximately comprised of 78% nitrogen, 21% oxygen, and 1% of various gases. We performed atmospheric measurements to determine the effect from the dominant species. Under laser exposure, nitrogen does not appear to have any detectable effect (Figure S6), while oxygen induces the characteristic change (Figure S7), suggesting oxygen is required for the mechanism. This is in agreement with previous work on $Bi_2Se_3$-$MoS_2$ 2D heterostructures.[39] Additionally, previous work found that oxygen in ambient is able to diffuse into $Bi_2Se_3$, and that is able to disrupt the surface crystal structure, suggesting a unique relationship exists between the two.[40]

Fick's 2nd Law of Diffusion fit both the PL intensity and peak position curves well, suggesting both values are closely coupled to the concentration of absorbed oxygen. The 2nd air PL intensity curve follows Fick's 2nd Law of Diffusion fit until ~150s (Figure 4k), but deviates after, following a linear behavior. This discrepancy is likely due to either residual $Bi_2Se_3$ remaining from the 1st air exposure, or a slight laser-spot shift as a result of atmosphere exchanges. The diffusion coefficients for the 2nd vacuum and 2nd air PL intensity were found to be 2.46 nm$^2$/s and 0.180 nm$^2$/s, respectively, while the peak positions were found to be 0.171 nm$^2$/s and 1.04 nm$^2$/s, respectively. Our findings suggest oxygen desorbs faster in vacuum compared to absorption in ambient, demonstrating hysteretic behavior. Additionally, we show a correlation between laser power and rate of diffusion, where higher laser powers facilitate faster diffusion (Figure S14). Further work is needed to elucidate changes to the diffusion rate as a function of atmosphere, temperature, and number of cycles. The curves are smooth, deterministic, and low noise, suggesting multiple states are accessible, and that each pixel can store numerous bits of information for superdense multistate multidimensional data storage. The changes are long-term stable in a $N_2$ atmosphere at room temperature for over 334 days (Figure S8), an attractive feature for applications.

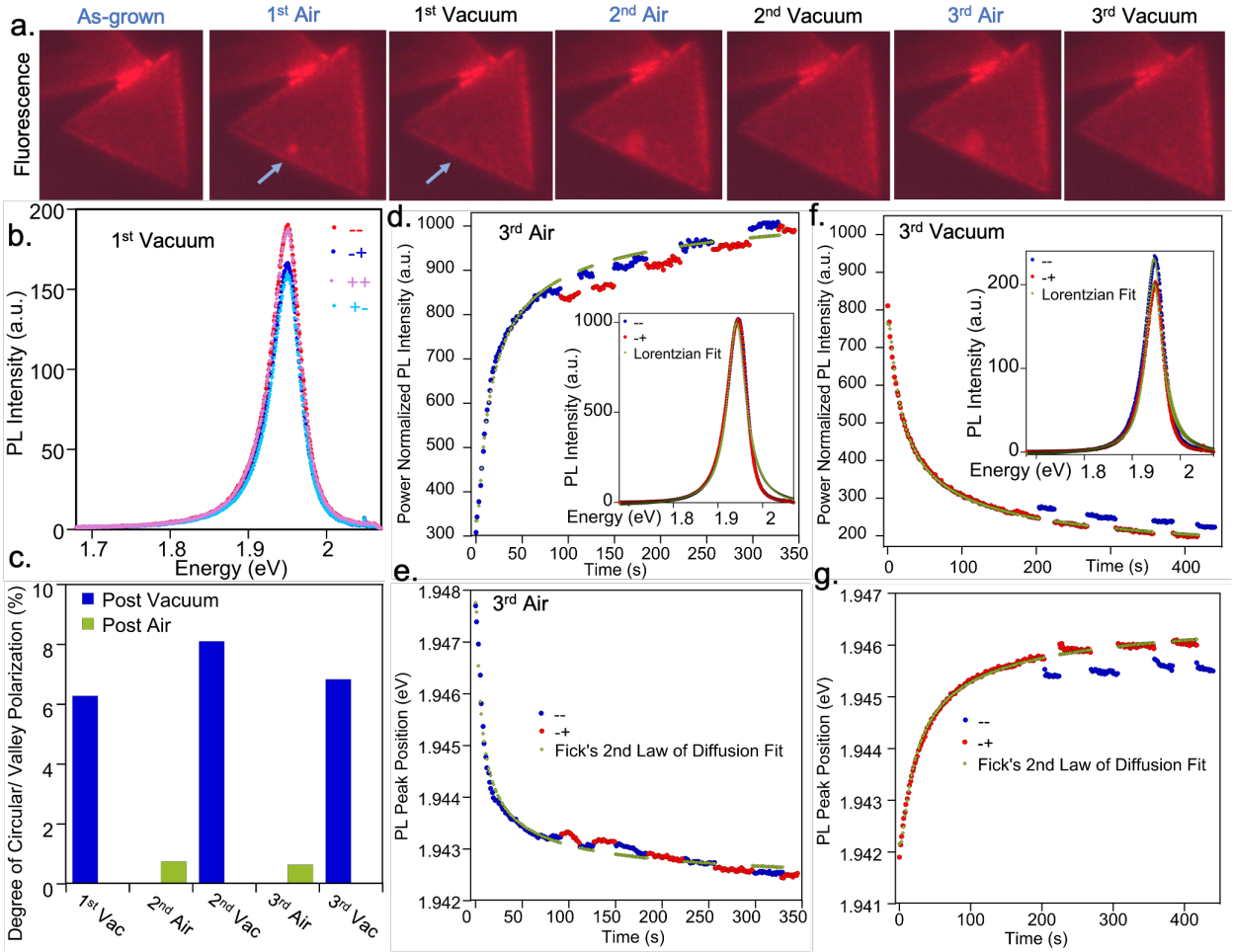

**Figure 5: DoCP and valley polarization can be reversed and modulated within a range.** (a) Fluorescence images after each laser-atmosphere exposure. (b) Both σ+ and σ- circular polarizations were probed, demonstrating that both valleys can be tuned and accessed. (c) DoCP is reversible and can be reliably tuned within a range. (d-g) Evolution of the PL intensity and peak position during 3rd air and 3rd vacuum. As expected, the PL intensity (peak position) increases (decreases) in air and decreases (increases) in vacuum. The behavior follows Fick's 2nd Law of Diffusion closely. As the response plateaued, both σ-- and σ-+ configurations were probed, allowing the DoCP to be extracted. Insets show final PL spectra with Lorentzian fits. More work is need to determine whether the difference in peak positions between σ-- and σ-+ is real or due to error.

Figure 5 demonstrates that, along with the PL intensity and peak position, changes to the DoCP can be reversed and reliably modulated within a range. Fluorescence images after each laser-atmosphere exposure (Figure 5a), show the PL intensity can be modulated between "bright" and "dark" states with spatial selectivity. Figure 5b show spectra from σ++, σ+-, σ--, and σ-+, demonstrating that both valleys can be accessed.

Figure 5c shows the measured DoCP, calculated using Eq. (1), after the last five laser-atmosphere exposures. The DoCP was reliably tuned between ~0.6% and ~7% after air and vacuum, respectively, a sizeable range considering that the intensity and peak position shifts are comparatively small. More specifically, consistent with our observations in Figure 3, previous work found an anticorrelation between PL intensity and DoCP, where greater intensity corresponded to higher DoCP.[23–26] Exciton lifetime likely plays a role (i.e., greater exciton lifetime corresponds to lower DoCP),[23–26] allowing more time for both a radiative transition or an intervalley scattering event.

Figure 5d-5g shows the PL intensity and peak position were tuned across comparatively narrow ranges (factor ×5 and 6meV, respectively). In contrast, Figure 3 demonstrates much larger ranges (factor ×161 and 38.4meV, respectively), suggesting much larger reversible DoCP shifts are feasible, and likely sample dependent. We believe sample response is influenced by $Bi_2Se_3$ stoichiometry, coverage, and layer thickness, along with TMD quality, suggesting material engineering is essential to increase the range of tunable and reversible DoCP.

Figure 5d-5g demonstrate the expected PL behavior, where the PL intensity and peak position increased (decreased) and decreased (increased) in air (vacuum), respectively. All PL behavior follows Fick's 2nd Law of Diffusion closely (Eq. 2), suggesting oxygen is being absorbed and desorbed. Due to equipment limitations, both the σ-- and σ-+ cannot be collected at the same time. To overcome this, the measurement system was switched between σ-+ and σ-- after the curves plateaued. Qualitatively a change in DoCP can be observed when examining the relative intensities of the σ-+ and σ-- curves. Each data point is extracted from robust fitting to a spectrum. Over 500 spectra were collected to produce the curves in Figure 5d-g, demonstrating the high control and low-error of the technology. The close fit to Fick's 2nd law of Diffusion suggests the system is deterministic, enabling high predictability and greater tunability. 1st Air, 1st Vacuum, 2nd Air, and 2nd Vacuum along with analysis are shown in Section S6 of the SI.

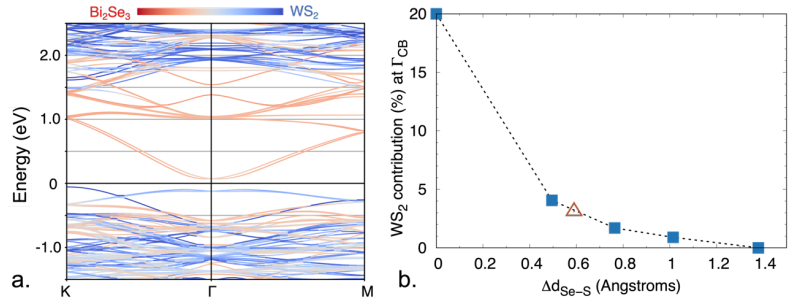

**Figure 6: Oxygen intercalation alters recombination pathways.** (a) Band structure of the $Bi_2Se_3$/$WS_2$ 2D heterostructure with spin-orbit coupling. The color of each state denotes the relative $WS_2$ (blue) versus $Bi_2Se_3$ (red) character. (b) Relative contribution of $WS_2$ to the CBM at $\Gamma$ as a function of the change in the interlayer separation distance with respect to the equilibrium separation distance, 3.221Å. The red triangle denotes the increase in $d_{Se-S}$ due to two $O_2$ molecules that are intercalated within the heterostructure.

We exposed bare monolayer $WS_2$ to a laser in ambient, but observed only comparatively small changes to the PL intensity and peak position, and DoCP (Section S5), demonstrating that both $Bi_2Se_3$ and oxygen diffusion are required to modulate the PL and valley polarization. In order to understand these changes, we perform first-principles calculations. The electronic structure of a monolayer $Bi_2Se_3$/monolayer $WS_2$ heterostructure calculated using density functional theory (DFT) calculations is illustrated in Figure 6.

The valence band maximum (VBM) of the heterostructure is a pure $WS_2$ state while the conduction band minimum (CBM) is hybridized with predominantly $Bi_2Se_3$ character and approximately 20% $WS_2$ character (Figure 6a). The band extrema at $\Gamma$ are comprised of out-of-plane $p_z$ orbitals of $Bi_2Se_3$ and $WS_2$ and therefore are sensitive to the minimum separation distance between Se and S, $d_{Se-S}$ which is 3.221Å in the equilibrium heterostructure.

If laser exposure in an $O_2$ environment leads to the intercalation of $O_2$ between $WS_2$ and $Bi_2Se_3$, this would increase $d_{Se-S}$, which would decrease the magnitude of the spatial overlap between the $p_z$ orbitals of S and Se. Indeed, when we optimize the atomic coordinates of the heterostructure with two $O_2$ molecules intercalated within the van-der-Waals interface of $Bi_2Se_3$-$WS_2$, we find $d_{Se-S}$ increases by 0.59 Å. In Figure 6b we plot the change in the degree of hybridization at $\Gamma$ as a function of increasing values of $d_{Se-S}$. It is clear that as the interlayer separation distance increases, the degree of hybridization decreases.

**Discussion**

Our calculations and proposed mechanism of $O_2$ intercalation imply the following. One process that is capable of causing changes in PL intensity is a nonradiative relaxation of photoexcited carriers from K. Prior to $O_2$ being intercalated, the hybridized CB state at Γ is a nonradiative pathway that photoexcited electrons at K can relax in to. If the contribution of $WS_2$ states to the CBM at Γ decreases, as we show can occur with intercalation, the probability for a nonradiative transition from a $WS_2$ derived state at K to Γ decreases. This would manifest in an increase in PL intensity. As $O_2$ de-intercalates, the interlayer separation decreases, the contribution of $WS_2$ to the CBM at Γ increases and the PL intensity decreases again. This is consistent with what we observe experimentally in Figure 4 and Figure 5.

Understanding the origins of the tunable DoCP is difficult. The combination of spin-orbit coupling and broken inversion symmetry in monolayer $WS_2$ leads to spin-valley locking at the K and K' valleys. The band extrema at K and K' can be independently accessed using CPL, enabling information processing schemes and other valleytronics applications. The excitons are able to steadily scatter to the opposite valley, which reduces the material's valley and circular polarization. Numerous intervalley scattering mechanisms have been proposed, which are discussed further in Section S7. Phonon-assisted scattering[15–18] and the Maialle-Silva-Sham (MSS) exchange interaction[16,19–21] are considered to be the most dominant mechanisms in monolayer TMDs.

We propose two ideas by which oxygen absorption and desorption in $Bi_2Se_3$ shifts the $WS_2$ DoCP. First, higher exciton lifetime correlates to lower DoCP.[23–26] Changes in the interlayer coupling strength between the $Bi_2Se_3$ and $WS_2$ may facilitate the presence (absence) of a non-radiative recombination pathway that decreases (increases) exciton lifetime. Second, the MSS mechanism, which requires the exchange interaction, is sensitive to the surrounding dielectric environment.[41] As oxygen modifies the $Bi_2Se_3$ crystal structure and the interlayer region, the dielectric environment is modified, thereby altering the intervalley scattering dynamics. Section S7 contains additional discussion and information.

We observe a PL peak position shift due to laser-air exposure, that is reversed through laser-vacuum exposure. After the 1st laser-oxygen exposure, both the PL intensity and peak position evolution follow Fick's 2nd Law of Diffusion, suggesting they are influenced by the same underlying mechanism. Monolayer TMDs are highly sensitive to their environment, and numerous factors affect the PL peak position,[42,43] due in part to tightly bound excitons, whose field lines extend outside the material. As the dielectric environment is modified, the exciton recombination energy and PL peak position can shift. One inconsistency is the observed difference in PL peak position between the σ- and σ+ configurations (Figure 5g). Although we suspect unidentified equipment error to be most likely, we cannot exclude the possibility that external effects from the $Bi_2Se_3$ and oxygen are asymmetrically modifying the *K* and *K'* valleys in $WS_2$. Section S8 contains additional discussion and information.

**Conclusion**
We demonstrate a method to tune the PL intensity, peak position, circular polarization, and valley polarization with high spatial selectivity using a laser in ambient on a $Bi_2Se_3$-$WS_2$ 2D heterostructure. The PL and DoCP can be reliably manipulated within a range, allowing the material to be erased and reused. Experiment and first-principles calculations suggest oxygen diffusion modulates the excitonic recombination pathways, where intercalation (deintercalation), increases (decreases) the probability of radiative recombination. The close fit to Fick's 2nd law of Diffusion suggests the system is deterministic, enabling high predictability and greater tunability. We demonstrate the possibility for high-resolution and precise tuning of submicron pixels for a variety of technologies, including multistate multidimensional optical data storage, valleytronics, and quantum communications.

**Methods**

**Material Growth:** Monolayer $WS_2$ is synthesized at ambient pressure in 2-inch diameter quartz tube furnaces on $SiO_2$/Si substrates (275 nm thickness of $SiO_2$). Prior to use, all $SiO_2$/Si substrates are cleaned in acetone, IPA, and Piranha etch ($H_2SO_4$+$H_2O_2$) then thoroughly rinsed in DI water. At the center of the furnace is positioned a quartz boat containing ~1g of $WO_3$ powder. Two $SiO_2$/Si wafers are positioned face-down, directly above the oxide precursor. A separate quartz boat containing sulfur powder is placed upstream, outside the furnace-heating zone, for the synthesis of $WS_2$. The upstream $SiO_2$/Si wafer contains perylene-3,4,9,10-tetracarboxylic acid tetrapotassium salt (PTAS) seeding molecules, while the downstream substrate is untreated. The hexagonal PTAS molecules are carried downstream to the untreated substrate and promote lateral growth of the monolayer $WS_2$. Pure argon (65 sccm) is used as the furnace heats to the target temperature. Upon reaching the target temperature of 825 °C, 10 sccm $H_2$ is added to the Ar flow and maintained throughout the 10-minute soak and subsequent cooling to room temperature.

$Bi_2Se_3$ was grown on top of the monolayer $WS_2$ using chemical vapor deposition (CVD) in a two-zone furnace with a 2" quartz tube. High-purity $Bi_2Se_3$ flakes are ground using a mortar and pestle into a fine dust. The powdered $Bi_2Se_3$ is placed in a ceramic boat and inserted into the furnace's quartz tube, and pushed into the center of the furnace's first zone. The monolayer TMD, which is on an $SiO_2$ substrate, is placed downstream of the $Bi_2Se_3$ into the center of the furnace's second zone. The furnace is pumped down to ~20mTorr. An argon (Ar) carrier gas is flown into the furnace at 80sccm. The $Bi_2Se_3$ is heated to 520°C, and the $WS_2$ are heated to 210°C. The ramp rate is ~55°C/min, and the total growth is 27 min.

**Transmission Electron Microscopy**
$Bi_2Se_3$-$WS_2$ 2D heterostructures were transferred onto a holey amorphous $SiN_x$ TEM grid using the water-assisted-pick-up transfer method.[44] Bright-field imaging, high-resolution TEM (HRTEM) and selected area electron diffraction (SAED) were performed with a JEOL JEM2200FS operating at 200 kV, equipped with a high-speed Gatan OneView camera. The HRTEM images were acquired at selected regions-of-interest as rapidly as possible to minimize sample alteration from the electron beam. The SAED patterns were internally calibrated to the WS2, and an aperture of approximately 200nm was used.

**Atmospheric measurements:** Atmospheric measurements were conducted inside a Janis ST-500 with vacuum, air, $N_2$ and $O_2$ connections. Figure 4 applied a 38.8μW 532nm laser without interruption at a 50x long working distance objective (1.65μW/μm$^2$). Spectra were measured at various intervals with a 0.8 sec. collection time. Figure 5 and Section S6 applied 84.7 μW 588nm laser without interruption at a 50x long working distance objective. Spectra were measured at various intervals with a 0.4 sec. collection time.

**Computational analysis and fitting:** All code as written in Python using the Spyder integrated development environment (IDE). Spyder belongs to the MIT License and is distributed through the Anaconda environment. Each spectrum is time stamped, and our algorithm extracted this info. The curve_fit() function with a variety of initial values and boundary conditions were used to verify fit robustness. The 1-σ uncertainty (error) was obtained from the curve_fit() function around the chi-squared minima for the applicable variables, and the error was propagated as appropriate to produce the 2-sigma (95.4% confidence) value shown.[37] The fitting and error values were corroborated using cross-validation, where we uniformly removed 20% of the data points from a spectrum and repeated fitting. This was completed five iterations on each spectrum removing a different 20% each time. No notable changes to the fitting were detected, suggesting noise is not skewing the fit.

**Laser exposure recipes in ambient:** Figure 2 applied a low-power (0.762µW) 532nm laser with a 50x objective (0.050µW/µm²) for 620s. A spectra were measured every 2 sec. for a 1 sec. collection time.

**Intermittent Laser Exposure Recipe for circular polarization data collection:** The σ++ and σ+- spectra are collected at a significantly diminished laser intensity (0.899µW) σ+ circular polarized 588nm laser for 10s each using a 100x objective (0.083µW/µm²). The very low laser power *at 588nm* laser does not appear to modify the sample. The laser intensity is then increased ×15 to 13.5µW for ~125s to controllably modify the 2D heterostructure. Such a recipe ensures that acquired σ++ and σ+- spectra accurately describe the system, and that the 2D heterostructure is not modified during data collection.

**Measuring and Removing Circular Polarization System Error:** To collect the different circular polarization configurations, the system rotates quarter and half wave plates, which have different transmission profiles, leading to error. To correct for intensity error, we collected numerous spectra of a GaAs sample in all circular polarization configurations (i.e., ++, +-, --, and -+). We made the ansatz that there should be no circular polarization in GaAs, and define the error as the computed the difference. To correct for peak position error, we collected numerous spectra of an as-grown $WS_2$ sample. We made the ansatz that the $WS_2$ peak position between – and -+ should be the same, and define the error as the computed the difference. More work is need to determine whether the difference in peak positions between σ-- and σ-+ is real or due to error. We do not imply that any difference between the σ-- and σ-+ peaks is necessarily physically significant.

**Density Functional Theory Calculations:** Our first-principles calculations are based on density functional theory within the projector-augmented wave method[45] as implemented in the VASP code.[46,47] We use the generalized gradient approximation defined by the Perdew-Burke-Ernzerhof (PBE) functional.[48] To determine the properties of the $Bi_2Se_3$/$WS_2$ heterostructure we use a 3x3x1 supercell of a single quintuple layer of $Bi_2Se_3$ and a 4x4x1 supercell of a single monolayer of $WS_2$ and strain the in-plane lattice parameters of $Bi_2Se_3$ to match the in-plane lattice parameters of $WS_2$. We treat the 6s, 6p electrons in Bi, 6s, 5d electrons in W, 3s,3p electrons in S and 4s,4p electrons in Se as valence in our PAW calculations. For the heterostructure calculations we use a (4x4x1) k-point grid when optimizing the structure and a (6x6x1) k—point grid to obtain total energies and electronic structure. We use a vacuum spacing of 20Å along the c-axis to avoid spurious interactions with periodically repeated surfaces. All of our calculations of the electronic structure include spin-orbit coupling. The atomic positions of the heterostructure were optimized with the Grimme-D3 van der Waals correction[49] with a force convergence criteria of 5 meV/Å.

We note that the PBE functional we use leads to an underestimated band gap for monolayer WS2 and likely underestimates the band gap of single quintuple layer of $Bi_2Se_3$. We performed hybrid functional calculations using the Heyd-Scuseria-Ernzerhof HSE06 functional[50] to verify that our calculations on the degree of hybridization are not impacted by the magnitude of the band gap in our calculation.

**Supporting Information**
AFM Scans (Section S1); Visual Demonstration of Robust Fit Quality (Section S2); Evolution of PL and Raman under laser exposure in different atmospheres (Section S3); Long Term Stability (Section S4); Control Experiments with monolayer $WS_2$ (Section S5); Demonstrating Reversibility of DoCP and valley polarization (Section S6); Amplifying Information on possible origin of tunable DoCP (Section S7).

# Supporting Information

**Laser-patterned submicron Bi$_2$Se$_3$-WS$_2$ pixels with tunable circular polarization at room temperature**

Zachariah Hennighausen[1,*], Darshana Wickramaratne[2], Kathleen M. McCreary[2], Bethany M. Hudak[2], Todd Brintlinger[2], Hsun-Jen Chuang[3], Mehmet A. Noyan[2], Berend T. Jonker[2], Rhonda M. Stroud[2], and Olaf M. van 't Erve[2,*]

[1] NRC Postdoc Residing at the Materials Science and Technology Division, United States Naval Research Laboratory, Washington, D.C. 20375, USA
[2] Materials Science and Technology Division, United States Naval Research Laboratory, Washington, D.C. 20375, USA
[3] Nova Research, Inc., Alexandria, VA 22308, USA

* Authors for correspondence, E-mail: zachariah.hennighausen.ctr@mail.nrl.navy.mil; olaf.vanterve@nrl.navy.mil;

*Keywords: circular polarization, TMD, 2D material*

## S1. AFM Scans: Bi$_2$Se$_3$ Layer count and growth morphology

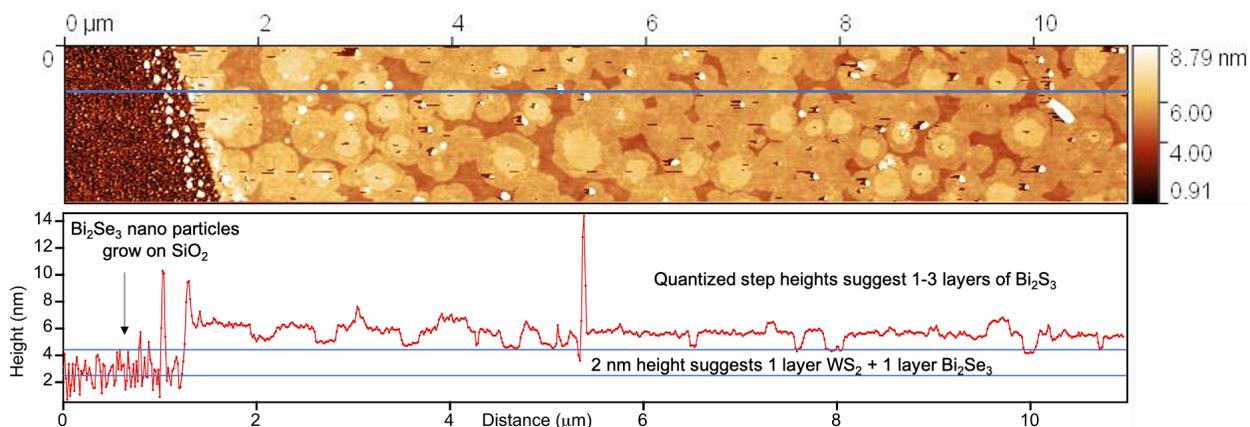

**Figure S1. Non-contact mode AFM scan of representative sample.** (top) AFM scan. (bottom) Line profile corresponding to inset blue line. Bi$_2$Se$_3$ nano particles grow on the SiO$_2$. Quantized step heights are observed, suggesting individual layers of Bi$_2$Se$_3$ are grown. The results indicate mostly bilayer Bi$_2$Se$_3$ was grown on monolayer WS$_2$, with gaps of monolayer Bi$_2$Se$_3$ and islands of trilayer Bi$_2$Se$_3$. The total scan was large (20μm×20μm) leading to bending at the edges. The above image is cropped from the larger scan to better display the data.

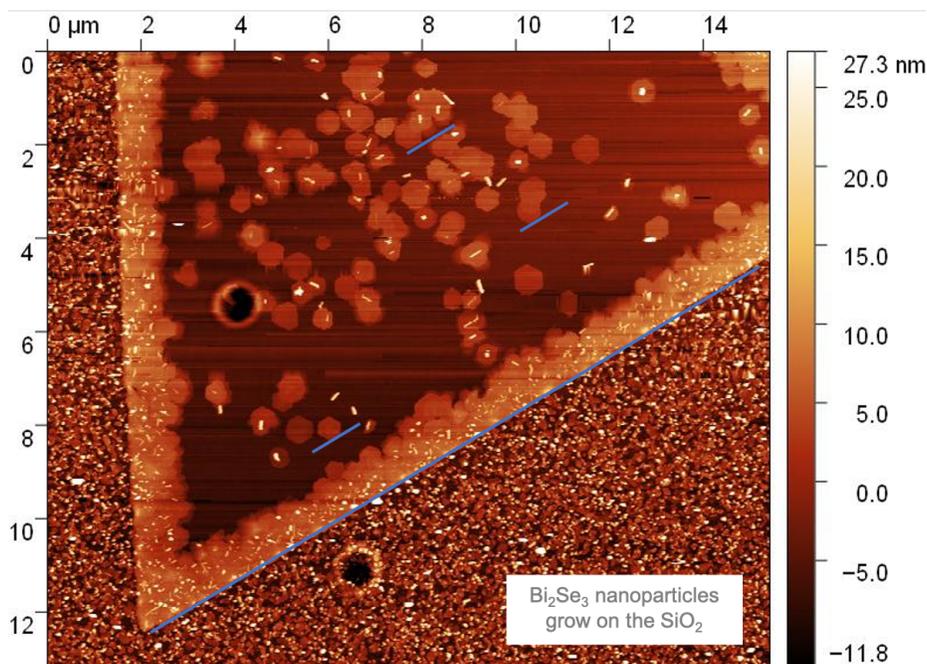

**Figure S2. AFM scan of incomplete growth showing crystalline Bi$_2$Se$_3$ on a WS$_2$ triangle.** The Bi$_2$Se$_3$ grows as well-defined hexagonal or triangular shapes, suggesting it grows crystalline. The edges of the Bi$_2$Se$_3$ hexagons qualitatively appear to be parallel to the WS$_2$ triangle, suggesting the Bi$_2$Se$_3$ grows at a 0° twist angle relative to the WS$_2$, in agreement with TEM results shown in Figure 1. Blue lines are added as a guide to the eye.

## S2. Visual Demonstration of Robust Fit Quality

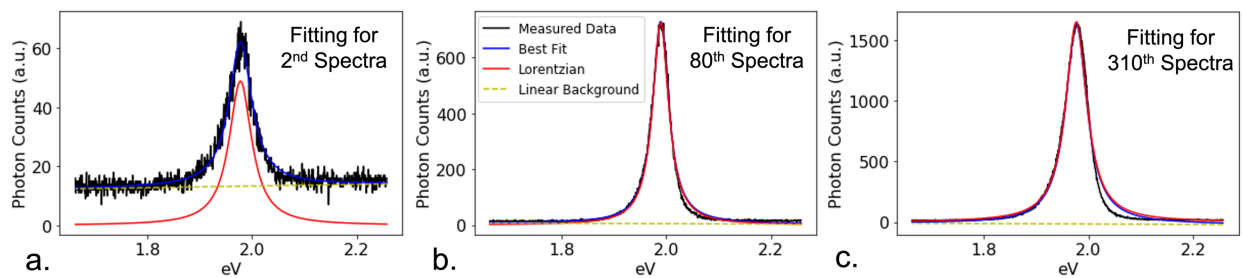

**Figure S3. Representative Lorentzian fitting for quantitatively extracting the PL intensity and peak position with low error.** (a)-(c) Representative spectra with Lorentzian fits from the session shown in Figure 2, where the sample was exposed to a laser in ambient. The PL intensity and peak position are quantitively extracted with low error by fitting them with a Lorentzian function and a linear background. All fitting is done using Python software, where the search algorithm is made sufficiently robust to ensure the global minimum is found with low error.[1] The method consistently produces a good fit across different data sets and samples, suggesting it sufficiently captures the spectrum. The 2-sigma (95.4%) confidence is calculated for all the fitting parameters, including the PL intensity and peak position. A gap is present at higher intensities on the right side of the spectra, possibly due to phonon side bands.[2]

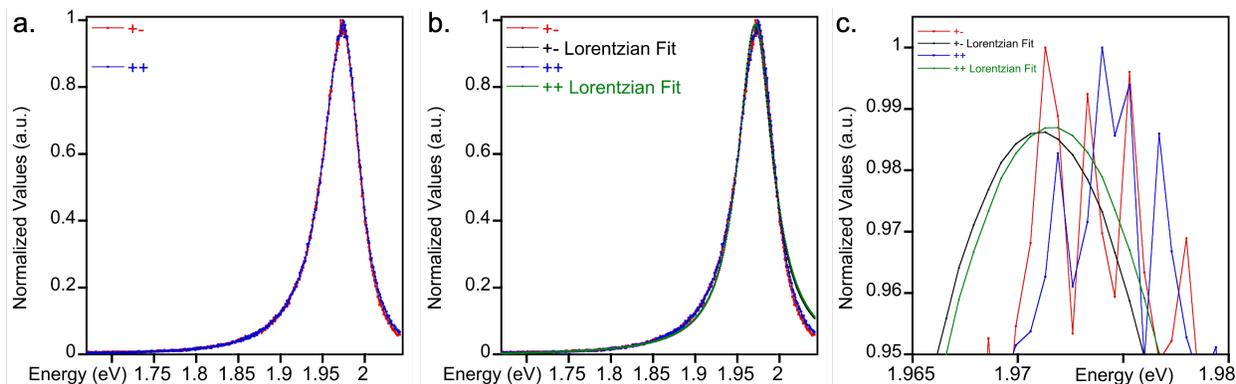

**Figure S4. Normalized σ++ and σ+- Spectra with Lorentzian fitting.** (a)-(b) Normalized spectra from the 33rd data point shown in Figure 3 without and with Lorentzian fits. (b) The Lorentzian fits do not capture the spectra width perfectly, possibly due to material effects. For example, phonon sidebands in monolayer TMDs have been shown to skew Lorentzian fits.[2] (c) Close-up of the peaks in (b) showing the fits are shifted from the experimental peak by ~4meV. More work is need to determine whether the difference in peak positions between σ-- and σ-+ is real or due to error. We do not imply that any difference between the σ-- and σ-+ peaks is necessarily physically significant.

## S3. Evolution of PL and Raman under laser exposure in different atmospheres (air, vacuum, N₂, and O₂)

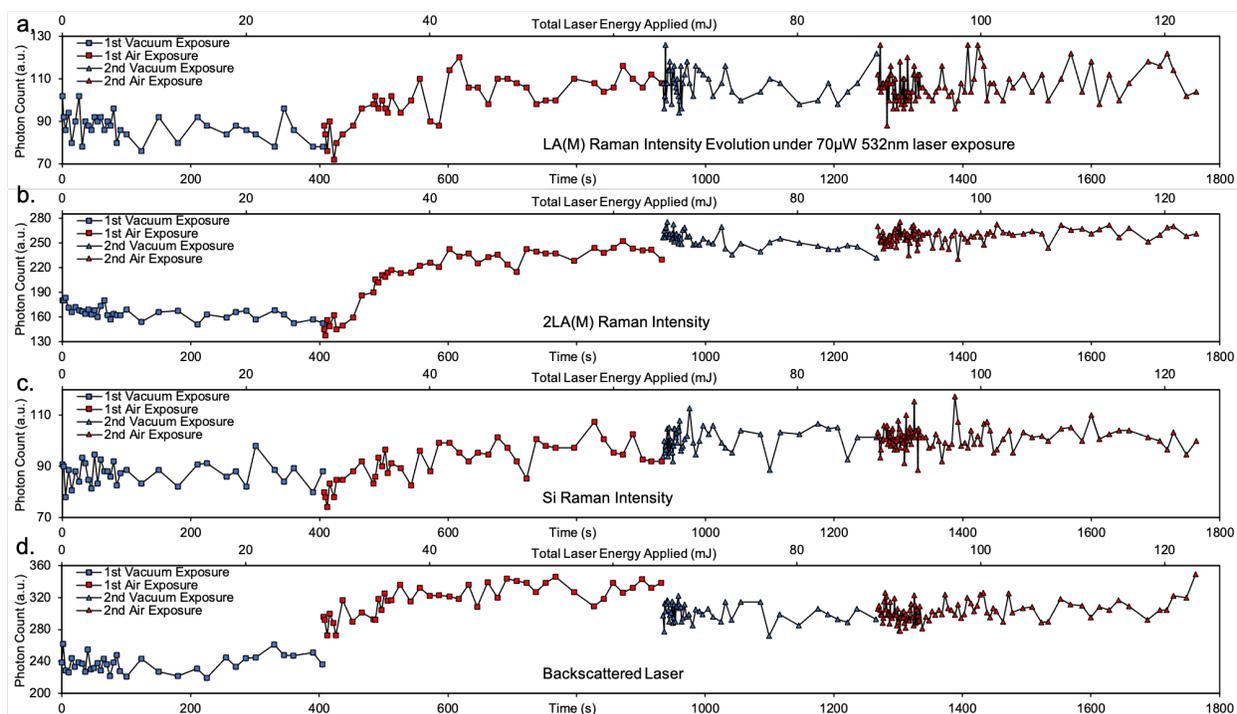

**Figure S5. No detectable effect of laser-vacuum exposure on the Raman modes.** (a)-(d) Although laser exposure in vacuum reverses the changes to the PL intensity and peak position, no changes to the Raman modes were detected. The above data is taken concurrent with the atmosphere cycling shown in Figure 4 of the main article. (a) LA(M), (b) 2LA(M), (c) Si, and (d) Backscattered laser.

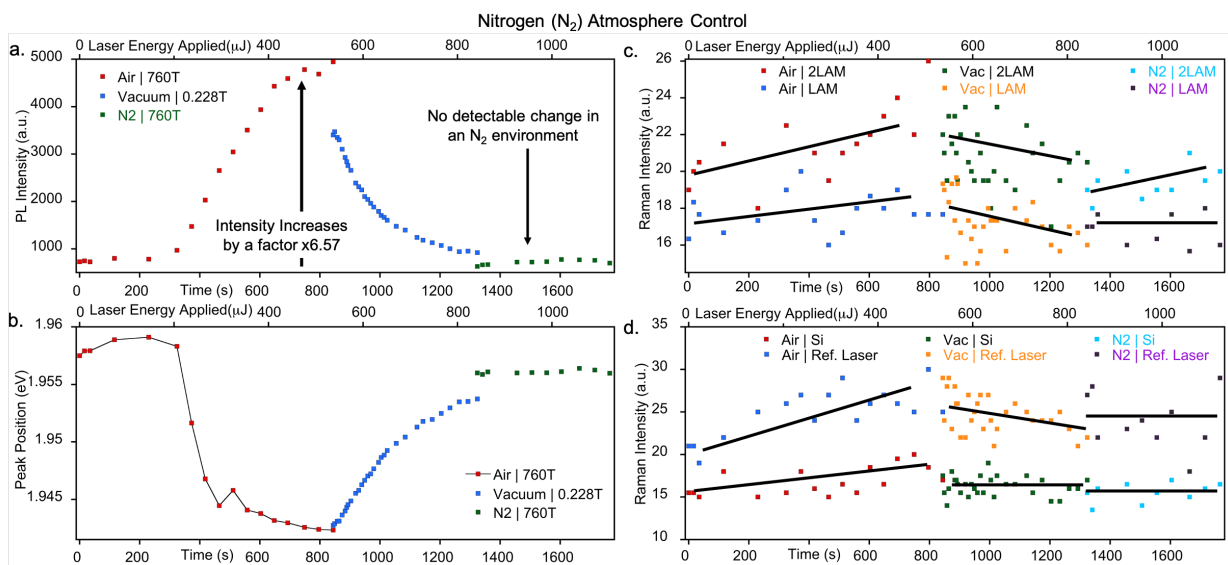

**Figure S6. No detectable change in nitrogen (N₂) atmosphere under laser exposure.** Evolution of PL (a) intensity and (b) peak position as a function of laser exposure in air (760T), vacuum (0.226T), and 99.9% nitrogen (N₂) (760T) atmospheres. We showed that changes are induced in a 99.99% oxygen environment (Figure S4); however, questions remained whether oxygen was simply diffusing into the 2D

heterostructure, and other gas species could induce this effect. No significant change is detected when the 2D heterostructure is exposed to a laser in a nitrogen environment, suggesting nitrogen is not diffusing into the 2D heterostructure, and/or their interaction has a negligible effect. Since $N_2$ is smaller than $O_2$, the results suggest that the small size of $O_2$ is not the primary mechanism, but that there is a chemical mechanism that facilitates the changes. When exposed to air and vacuum, the expected changes are induced, demonstrating that the material functioned as expected. The 2D heterostructure is exposed to a low-power (38.8µW) 532nm laser at a 50x long working distance objective (1.65µW/µm²). (c)-(d) LA(M), 2LA(M), Si, and backscattered signals, respectively. The resolution is too low to extract notable insight, but the signals in air appear to evolve in agreement with Figure S5, while the signals in vacuum appear to decrease.

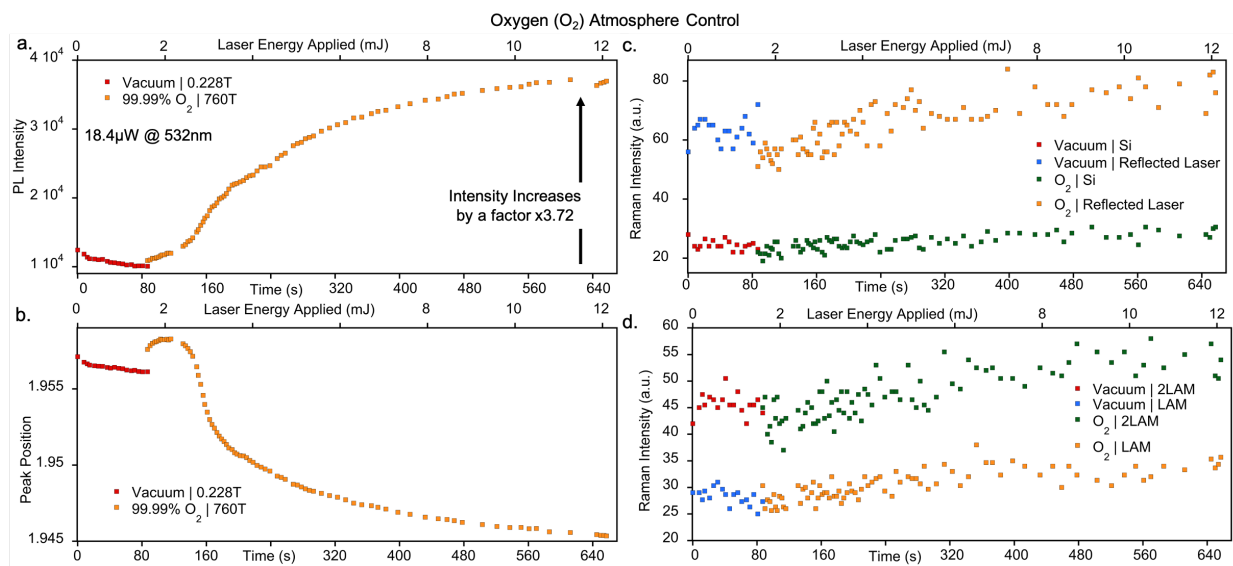

**Figure S7. Oxygen is required to induce changes under laser exposure.** Evolution of PL (a) intensity and (b) peak position as a function of laser exposure in vacuum (0.226T) vs. 99.9% oxygen (760T) environments. The 2D heterostructure is exposed to a low-power (18.4µW) 532nm laser at a 50x long working distance objective (0.780µW/µm²). When exposed to oxygen, changes are induced, suggesting oxygen is required for the primary mechanism. Changes are not induced when exposed to nitrogen while a laser is applied. (c)-(d) LA(M), 2LA(M), Si, and backscattered signals, respectively. The signals in air appear evolve in agreement with Figure S5.

## S4. Long term stability in N$_2$ at room temperature

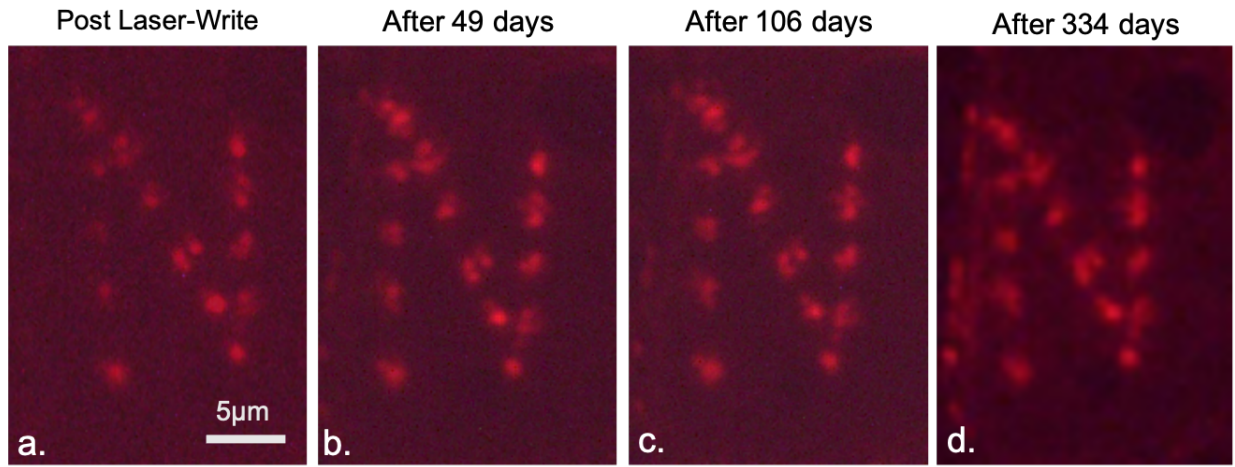

**Figure S8. Changes are stable for long term with no detectable degradation.** (a)-(d) 2D Heterostructures were written with a laser, and kept in a nitrogen environment at room temperature. No notable changes to the laser exposed were observed, suggesting the changes are robust and stable. (a) post laser-write. (b) after 49 days. (c) after 106 days. (d) after 334 days.

## S5. Control experiments: Monolayer WS$_2$ response (without Bi$_2$Se$_3$)

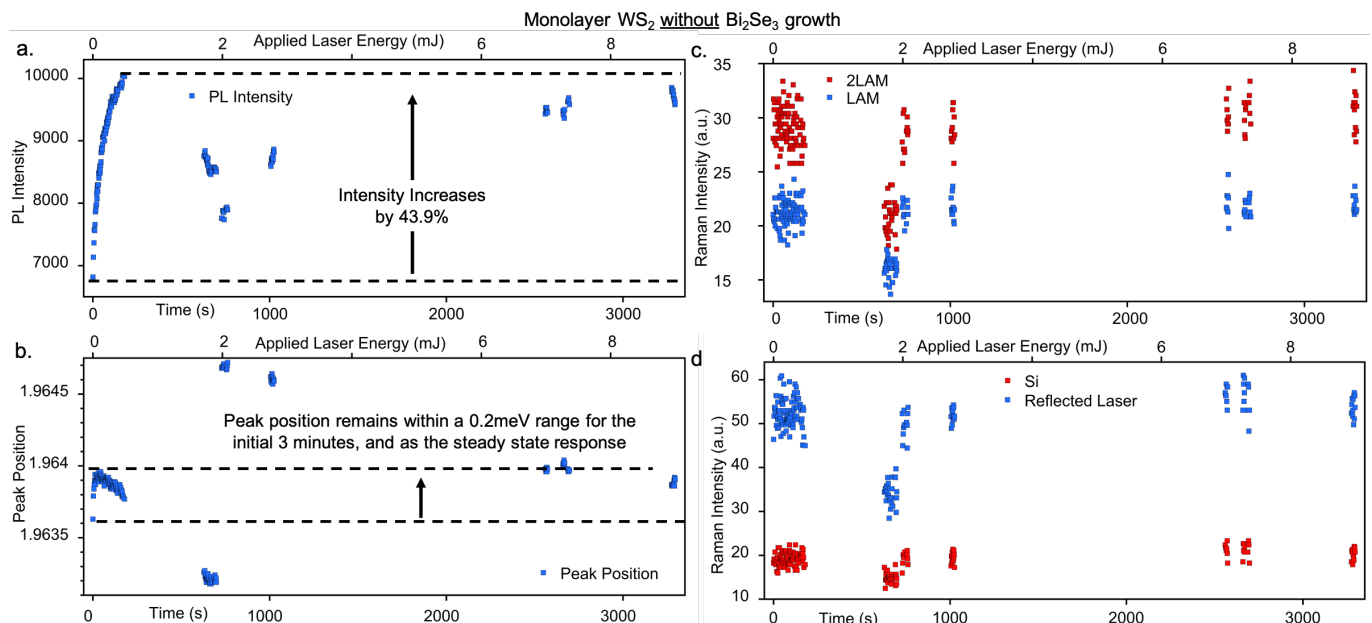

**Figure S9. Evolution of monolayer WS$_2$ (without Bi$_2$Se$_3$) PL and Raman spectra under 532nm laser exposure in ambient.** Monolayer WS$_2$ was grown using the CVD recipe described in the methods, and then subject to equivalent Bi$_2$Se$_3$ growth conditions. More specifically, the WS$_2$ was placed in a vacuum environment with Bi$_2$Se$_3$; however, the WS$_2$ was heated to above the Bi$_2$Se$_3$ deposition temperature, thereby preventing deposition of Bi$_2$Se$_3$. Evolution of the (a) PL intensity and (b) peak position subject to a 2.73µW 532nm laser at 100x objective (0.92µW/µm²) for 55min. Changes to the PL intensity and peak position are much smaller than what is observed in the Bi$_2$Se$_3$-WS$_2$ 2D heterostructure, suggesting the effects demonstrated in this work are overwhelmingly influenced by the Bi$_2$Se$_3$. Over the course of the 55min, the PL intensity and peak position fluctuate, suggesting the WS$_2$ is being modified. (c)-(d) Raman peaks and the backscattered laser peak detected concurrent with the PL spectra. None of the changes to the PL or Raman spectra are as large as those observed in the Bi$_2$Se$_3$-WS$_2$ 2D heterostructure, suggesting that when the Bi$_2$Se$_3$ becomes more transparent, more light will enter WS$_2$.

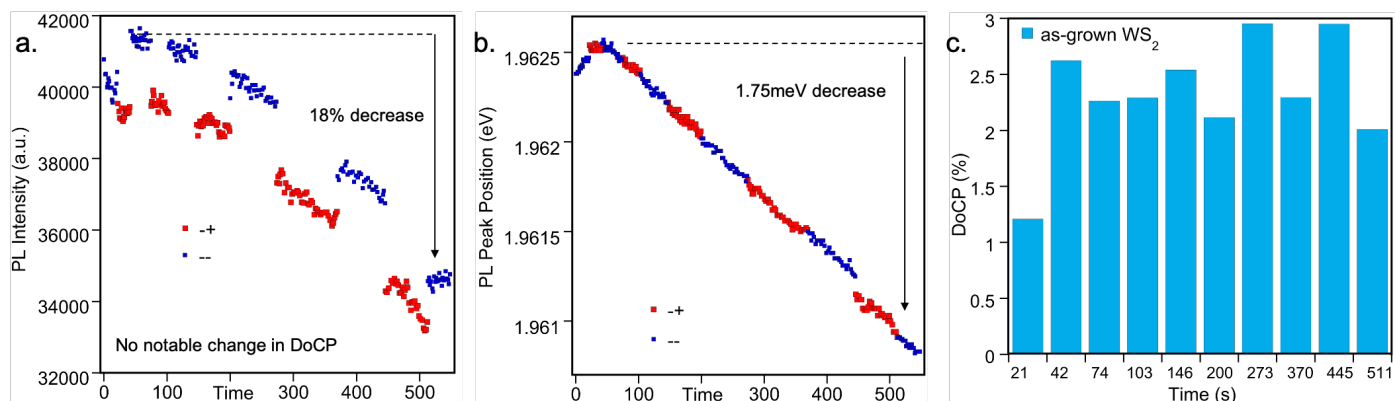

**Figure S10. Evolution of as-grown monolayer WS$_2$ (without Bi$_2$Se$_3$) under 588nm laser exposure in ambient.** Monolayer WS$_2$ was grown using the CVD recipe described in the methods. Evolution of the (a) PL intensity and (b) peak position subject to a 84.7µW 588nm laser at 50xL objective for 10min. Both – and -+ CP directions were probed. Changes to the PL intensity and peak position are much smaller than what is observed in the Bi$_2$Se$_3$-WS$_2$ 2D heterostructure. The σ-+ peak position data was shifted upward 0.33meV to account for equipment error (see methods). (c) DoCP as a function of time. DoCP fluctuates within 0.94% after the initial exposure, suggesting the effects demonstrated in this work are overwhelmingly influenced by the Bi$_2$Se$_3$.

## S6. Demonstrating reversibility of DoCP and valley polarization

We demonstrate that the DoCP can be reversibly tuned within a range. A sample was exposed to different laser wavelengths, powers, and polarizations in both air and vacuum atmospheres. Figure 11 below summarizes the experiment. We apply a 532nm laser in the 1st air and 1st vacuum exposures to modify the $Bi_2Se_3$, and a 588nm laser in all subsequent exposures to measure the DoCP. Results and analysis from each laser-atmosphere exposure are shown below in Figure 5 (main manuscript), and Figures S11-S13.

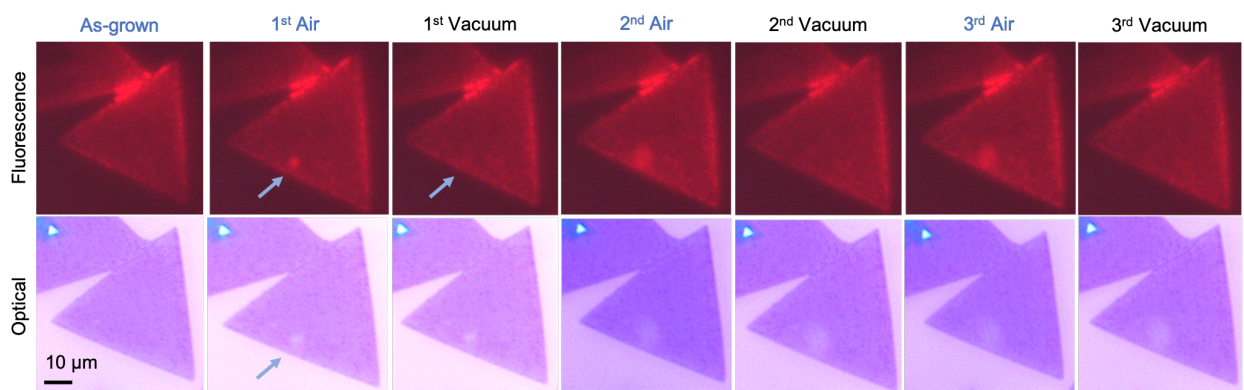

**Figure S11: Summary of experiment probing reversibility of DoCP in $Bi_2Se_3$-$WS_2$.** 532nm was applied in 1st air and 1st vacuum, followed by 588nm laser for remaining atmospheres. (Upper row) Fluorescence images showing the PL intensity can be modulated between bright and dark states. (Bottom row) Corresponding optical images, showing the laser-$O_2$ color change is not reversed.

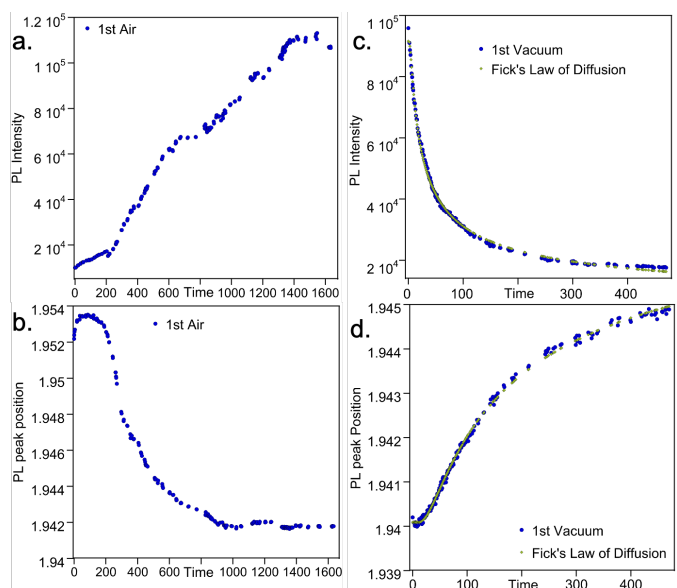

**Figure S12: Reversible DoCP - 1st air and 1st vacuum - PL intensity and peak position evolution.** Figure S12a-b show the expected PL intensity and peak position evolution during the 1st air exposure. The PL steadily increases with a non-linear feature in the middle, while the peak position increases before decreasing, in agreement with Figures 2, 4, and S7. Figure S12c-d show the expected PL intensity and peak position evolution from subsequent vacuum exposures. The intensity and peak position behavior follow Fick's 2nd Law of Diffusion closely. DoCP measurements were not collected while the 532nm laser was applied, since DoCP measurements are best measured using the 588nm laser. The 532nm laser, however, provides more control when modifying the Bi$_2$Se$_3$ during the 1st air exposure. More specifically, the 1st air exposure is much slower and less predictable compared to subsequent air exposures. We attribute this to the Bi$_2$Se$_3$ being modified, an assertion supported by Raman and optical measurements (see Figure 2 and Figure S5). Using the 532nm laser in our setup provides greater control over applied power, simultaneous measurement of the Raman spectra, and higher signal sensitivity.

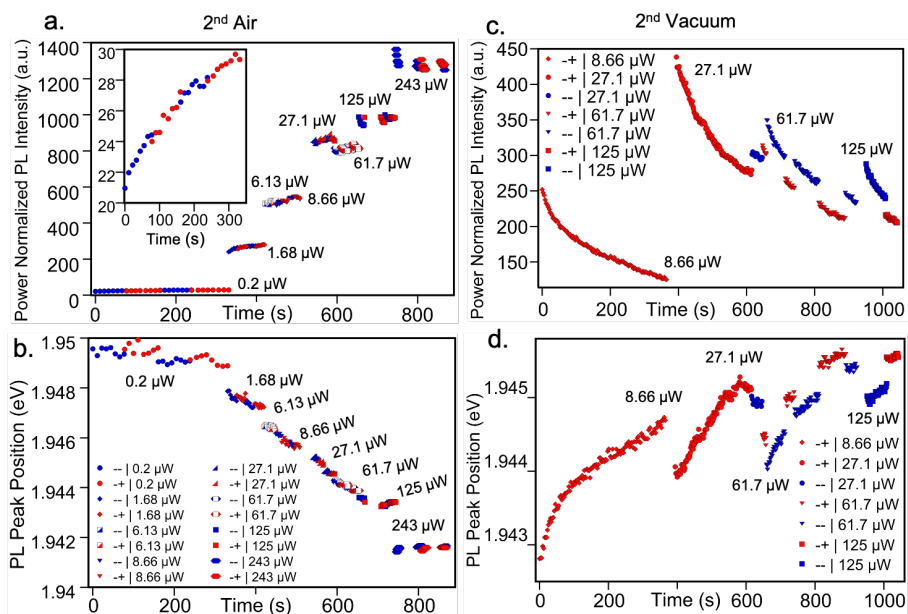

**Figure S13: Reversible DoCP – 2nd air and 2nd vacuum - PL intensity and peak position evolution.** The material was probed with a variety of powers in both the -+ and – configurations. (a-b) 2nd Air PL intensity and peak position. An evolution following Fick's 2nd Law of diffusion was not observed, likely due to over exposure. To collect the low-power data points with sufficient intensity, long collection times are required. We believe the Bi$_2$Se$_3$ saturated with oxygen during the initial data collection. (c-d) 2nd Vacuum PL intensity and peak position. As expected, the PL intensity decreases, peak position increases, and the curve shapes appear to follow Fick's 2nd Law of Diffusion. Together, this suggests oxygen is being desorbed. Note: The data is plotted with the power normalized, showing photons/(second*µW), thereby revealing exciton dynamics. More work is need to determine whether the difference in peak positions between σ-- and σ-+ is real or due to error. We do not imply that any difference between the σ-- and σ-+ peaks is necessarily physically significant.

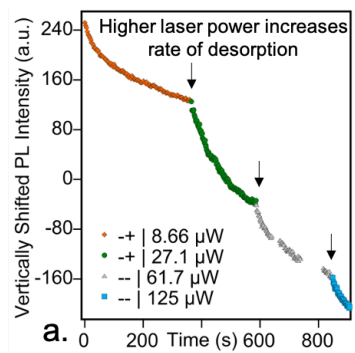

**Figure S14: Analysis of power measurements collected during 2nd air and 2nd vacuum.** (a) Power measurements taken during 2nd vacuum show that the desorption rate is coupled to the power applied. When the power was increased, the desorption rate increased as well.

## S7. Amplifying information on possible origin of tunable DoCP (valley polarization)

The combination of spin-orbit coupling and broken inversion symmetry in monolayer $WS_2$ leads to spin-valley locking at the *K* and *K*′ valleys. The band extrema at *K* and *K*′ can be independently accessed using circularly polarized light (CPL), enabling information processing schemes and other valleytronics applications. The excitons are able to steadily scatter to the opposite valley, which reduces the material's valley and circular polarization.

Numerous intervalley scattering mechanisms have been proposed as a source of the reduction of DoCP in monolayer TMDs, including D'yakonov-Perel' (DP),[3–5] Elliott-Yafet (EY),[3,5] electron-phonon and exciton-phonon interactions,[5–12] and the Maialle-Silva-Sham (MSS) mechanism (exchange interaction).[6,13–19] Below we very briefly review each mechanism, and then discuss how oxygen absorption into the $Bi_2Se_3$ might affect intervalley scattering.

The DP mechanism requires both time-reversal symmetry and broken inversion symmetry. As a result, the spin-orbit coupling produces a *k*-dependent effective magnetic field, which induces the spin of electrons at different momenta to precess in different directions, leading to spin relaxation.[3–5]

The EY mechanism requires a strong spin-orbit coupling, which leads to a mixing of the spin-up and spin-down states. An outcome is that spin-independent scattering events can lead to spin flips and consequent relaxation of the net spin polarization.[3–5]

In electron-phonon and exciton-phonon relaxation, a combination of phononic scattering, exchange interactions, and spin-orbit interactions, can lead to loss of both valley and spin-polarization in the excited electronic or excitonic population.[5–12]

The MSS mechanism requires an exchange interaction between excitons in opposite valleys. The exchange interaction, a many-body effect, produces an in-plane effective magnetic field. The magnetic field has a strong momentum dependence, and momenta averaging leads to decoherence (i.e., valley depolarization).[6,13–19] MSS mechanism can be visualized as a virtual recombination of an exciton in one valley, and the generation of a new exciton in the other valley. Since the center-of-mass momenta of the exciton is zero, the mechanism does not change the total momentum of an exciton, and momentum conservation is not an issue. The Coulomb force and screening electrons account for momenta conservation for trions. Interestingly, when an electron is scattered and its momentum shifted, the time-averaged momentum-dependent effective magnetic field experienced is reduced, thereby decreasing the decoherence time. Stated simply, scattering events can actually *increase* the valley polarization within the MSS framework.

It is an open research question how the mechanisms interact with another and what the dominant mechanisms are in monolayer TMDs. Based on previous research, we suspect the exciton-phonon and MSS mechanisms are the most active in our system.

Proposed Framework to understand how the oxygen concentration modulates DoCP:

We propose two ideas by which oxygen absorption and desorption in Bi$_2$Se$_3$ modifies the WS$_2$ DoCP and valley polarization. First, the MSS mechanism, which requires the exchange interaction, is sensitive to the surrounding dielectric environment. As oxygen is absorbed into Bi$_2$Se$_3$ and the interlayer region, the dielectric environment is modified, thereby altering the intervalley scattering dynamics. Second, exciton lifetime directly influences the DoCP. Changes in the interlayer coupling strength between the Bi$_2$Se$_3$ and WS$_2$ may facilitate the presence (absence) of a non-radiative recombination pathway that decreases (increases) exciton lifetime.

Amongst the various mechanisms that have been suggested to contribute to intervalley scattering in the TMDs, the MSS mechanism in particular is sensitive to the surrounding dielectric environment because the dielectric surrounding effectively screens the exchange interaction.[20,21] As oxygen is absorbed and desorbed from the Bi$_2$Se$_3$, the dielectric environment and effective screening of the exchange interaction is shifted proportional to the amount of oxygen present in Bi$_2$Se$_3$. This raises the possibility that the DoCP is shifted proportional to the concentration of oxygen within the Bi$_2$Se$_3$/WS$_2$ heterostructure. Further research is required to determine whether the dielectric changes are sufficient to shift the DoCP as observed experimentally.

Previous work demonstrated that valley polarization is dependent on exciton lifetime, where longer lifetimes correlate to smaller valley polarization.[22–24] The longer the exciton lifetime, the more time is available for it to scatter to the opposite valley. This is in line with what both the exciton-phonon and MSS mechanisms predict.

We speculate that when the Bi$_2$Se$_3$ and WS$_2$ are well-coupled (no oxygen), a non-radiative recombination channel is formed that decreases the exciton lifetime. When the interlayer coupling is diminished through oxygen absorption, the non-radiative channel is suppressed, and the exciton lifetime increases, thereby decreasing the valley polarization. This is in line with theory calculations in Figure 5, which predict hybridization between the Bi$_2$Se$_3$ and WS$_2$ that is dependent on the interlayer separation.

## S8. Amplifying information on possible differences between measurements of the σ- and σ+ configurations

Our findings suggest that the material has a comparatively stable PL, and the robust fitting is highly repeatable. As a result, we were able to detect a stable 0.41meV difference in peak position between the σ-- and σ-+ configurations (see Figure 5g). Although the difference is outside error and repeatable, it is a very small value considering that the FWHM is over 100x greater at 41.8meV. Considering how minuscule the difference is in context of the entire spectra, we are suspicious that we have identified and removed all necessary equipment error to observe such small changes with high confidence.

In Figure 5g, we are uncertain why there is a difference in peak position the between the σ-- and σ-+ configurations, but suspect equipment error as most likely. However, we cannot exclude the possibility that external effects from the $Bi_2Se_3$ and oxygen are asymmetrically modifying the $K$ and $K'$ valleys in $WS_2$.

In a perfect monolayer $WS_2$ crystal at 0°K, theory predicts that the band gap at $K$ and $K'$ are equivalent, suggesting there should be no changes in peak position from radiated spectra. Further, in our opinion, the monolayer TMD research community generally assumes the spectra should be identical (or nearly identical) between σ-- and σ-+ configurations of a bare monolayer $WS_2$ that is free from external effects. This ansatz, however, breaks down when the monolayer TMD is modified or subject to external effects. For example, high magnetic fields,[25] the magnetic proximity effect,[26] and $MoO_3$-coupling[27] have demonstrated the ability to produce large asymmetric responses from the $K$ and $K'$ valleys.

Regarding error effects, the equipment uses a spectrometer to measure the intensity at 1017 different, equally-spaced wavelengths. Despite our best efforts, we suspect a small amount of equipment error remains that is difficult to quantify independently at each wavelength, but can nonetheless have an effect on the fitting process because this captures the entire spectrum. More specifically, if background error slightly shifts the spectra tails, this would shift the fitting slightly.

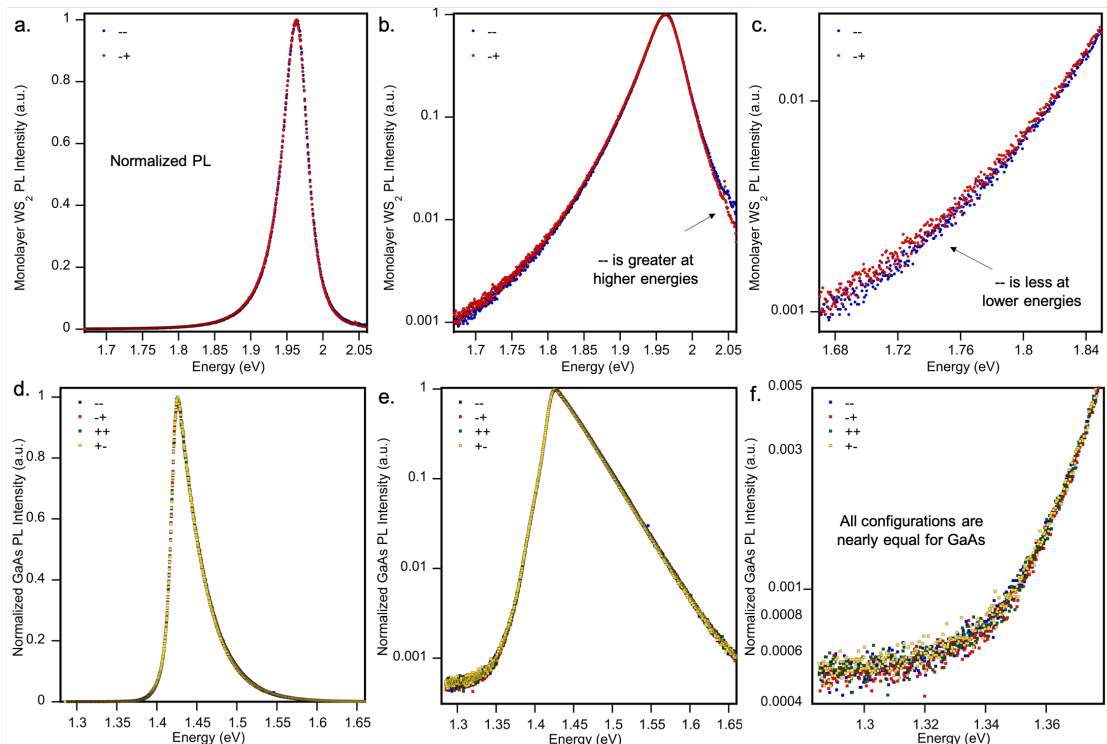

**Figure S15: Qualitative error analysis and equipment calibration data using as-grown monolayer WS$_2$ and GaAs control samples.** (a)-(c) As-grown monolayer WS$_2$ normalized PL intensity plots of σ-- and σ-+ configurations. (a) Qualitatively no obvious differences can be detected. (b)-(c) However, when plotted on a log scale, the tails of the spectra show slight changes, which influence fitting. The slight changes at the tails might be due to equipment error, Raman, or silicon effects. (d)-(f) GaAs control samples are used to calibrate the equipment. No obvious changes were detected between all the configurations. Of note, the GaAs spectra were collected at lower wavelengths.

Figure S15 shows qualitative error analysis and equipment calibration data using PL from as-grown monolayer WS$_2$ and GaAs control samples. Figure S15a-c show normalized WS$_2$ spectra from the σ-- and σ-+ configurations. Only when the data is viewed as a log plot can changes in the tails be observed, suggesting error is introduced from the equipment, silicon, or Raman signal. Figure S15d-f show normalized GaAs spectra, where no obvious changes can be observed.